\documentclass[%
 aip,
 amsmath,amssymb,
reprint,%
]{revtex4-1}
\usepackage{multirow}
\usepackage{graphicx}
\usepackage{dcolumn}
\usepackage{bm}
\usepackage{hyperref}
\usepackage[mathlines]{lineno}
\usepackage{xcolor}

\begin{document}

\title{Mechanisms of strength and hardening in austenitic 
stainless 310S steel: Nanoindentation experiments and multiscale modeling}                      

\author{F. J. Dom\'inguez-Guti\'errez}
\email[Corresponding author: ]{javier.dominguez@ncbj.gov.pl}
\author{K. Mulewska}
\affiliation{NOMATEN Centre of Excellence, National Centre for Nuclear Research, ul. A. Sołtana 7, 05-400 Otwock, Poland}

\author{A. Ustrzycka}
\affiliation{Institute of Fundamental Technological Research, , Polish Academy of Sciences, Warsaw, Poland}
 \author{R. Alvarez-Donado}
 \affiliation{NOMATEN Centre of Excellence, National Centre for Nuclear Research, ul. A. Sołtana 7, 05-400 Otwock, Poland}
 \author{A. Kosi\'nska}
 \affiliation{NOMATEN Centre of Excellence, National Centre for Nuclear Research, ul. A. Sołtana 7, 05-400 Otwock, Poland}

 \author{W. Y. Huo}
 \affiliation{NOMATEN Centre of Excellence, National Centre for Nuclear Research, ul. A. Sołtana 7, 05-400 Otwock, Poland}
\affiliation{College of Mechanical and Electrical Engineering, Nanjing Forestry University, Nanjing, 210037, China}

\author{L. Kurpaska}
\author{I. Jozwik}
 \author{S. Papanikolaou}
 \affiliation{NOMATEN Centre of Excellence, National Centre for Nuclear Research, ul. A. Sołtana 7, 05-400 Otwock, Poland}
 \author{M. Alava}
 \affiliation{NOMATEN Centre of Excellence, National Centre for Nuclear Research, ul. A. Sołtana 7, 05-400 Otwock, Poland}
  \affiliation{Department of Applied Physics, Aalto University, P.O. Box 11000, 00076 Aalto, Espoo, Finland}

\date{\today}

\begin{abstract}
Austenitic stainless steels with
low carbon have exceptional mechanical properties and are capable
to reduce embrittlement, due to 
high chromium and nickel alloying, thus they are very attractive for 
 efficient energy production in extreme environments. It is 
key to perform nanomechanical investigations of the role of chromium 
and the form of the particular alloy composition that give rise to the excellent mechanical properties of steel.
We perform nanoindentation experiments and molecular dynamics (MD) 
simulations of FCC austenitic stainless steel 
310S, using established interatomic potentials, and we use a comparison 
to the plastic behavior of NiFe solid solutions under similar 
conditions for the elucidation of key dislocation mechanisms. 
We combine EBSD images to connect 
crystalline orientations to nanoindentation results, and provide 
input data to MD simulations for modeling mechanisms
of defects nucleation and interactions. 
The maps of  impressions after nanoindentation indicate 
that the Ni-Fe-Cr composition in 310S steel leads to strain localization
and hardening.
A detailed analysis of the dislocation dynamics at different 
depths leads to the development of an experimentally consistent 
Kocks-Mecking-based continuum multiscale model. Furthermore, the
analysis of geometrically necessary dislocations (GND) shows to be
responsible for exceptional hardness at low depths, predicted by the 
Ma-Clarke's constitutive model.
\end{abstract}




\maketitle
\section{Introduction}
\label{sec:intro}

Stainless steels with promising thermo-mechanical properties
are currently proposed to be used in extreme environments 
applications such us new and efficient nuclear energy systems
or demanding transport and construction industries 
\cite{LO200939,CHEN201783,WEN201734,ISAKOV2016258}. 
310S stainless steel (SS) can also be used for thermal power
plants,
where containers present distortion during the process of 
being filled with high-level nuclear waste glass at elevated
temperature requiring a high strength of the material 
\cite{Zhang2019, Ustrzycka2012, Ustrzycka2013}. 
Finally, SS can also be used for manufacturing radiant
tubes, thermowells, burners and combustion chambers, and 
annealing covers \cite{Zhang2019}. 
Among these applications, nuclear reactor environment seems to 
be the most challenging. 
Despite the high demands related to structural, mechanical and 
thermal properties stability at reactor environment,  
austenitic SS is considered as excellent candidate for 
applications in Generation IV - Super-Critical Water Reactors 
(SCWR) due to its superior corrosion resistance. 
SCWR type nuclear reactors are designed to operate above the 
thermodynamic critical point of water (22.1 MPa, 374 C). 
According to the nuclear community, this type of reactor is 
supposed to be considered as one of the most promising future 
Generation IV concepts. 
This is due to its simplified design, compact layout, and high
thermal efficiency in comparison to standard PWR or LWR systems. 
However, one should remember that above the thermodynamic 
critical point, the coolant is more corrosive which is a
challenge for in-core structural components (especially fuel
cladding material). 
In addition, the so called hot-spots, 
which are typical for SCWR technology, locally may reach even 600 C. 
Beyond this temperature, strength decreases and the material may oxidize
rapidly due to the development of Cr$_{23}$C$_{6}$ particles
which are preferentially located at grain boundaries. 
This increases brittleness and further deteriorates corrosion
resistance. 
It is this temperature ($600^o$C) where there is significant mechanical 
properties drop, and a detailed understanding of plastic deformation
phenomena in SS is required before and 
after irradiation. 

It is known that corrosion resistance is provided by the 
presence of Chromium ($\sim$20\%) in the material. 
At the same time, high strength and ductility are maintained
due to the inclusion of high nickel content ($\sim$20\%) in
the material. 
Moreover, high Ni content makes this kind of steel to
not exhibit a strain-induced phase transformation \cite{LO200939,LiuGao}. 
Finally, additions of carbon in SS are limited and optimized
to reduce the prone to embrittlement and improve its creep
deformation resistance. Despite having a fairly good 
understanding of the effect of individual elements and their
concentration on the functional properties, the microscopic
mechanisms of plastic 
deformation are still an intriguing challenge. 
For this reason, in this work, experimental 
and numerical investigations are carried out towards understanding SS' 
themo-mechanical properties. 

It is well known that plasticity properties and physical
mechanisms associated with the deformation of a material can be 
estimated from nanoindentation data 
\cite{CHEN201783,XIA2019108169,SCHUH200632,Nix1998,oliver_pharr_1992} 
where hardness and flow stress are well explained in the 
literature \cite{Matucci2021, Cui_2020, Zhang2004, Zhang2006, 
Rodriguez2003}. Hardness is defined as the ability 
of a material to resist plastic deformation. Tabor’s relationship 
between hardness and yield strength \cite{Tabor_1956} is formulated 
in the following way  $\sigma_y \simeq \psi H$, where $\sigma_y$ is 
the tensile yield
strength, $H$ is the indentation hardness measured and $\Psi$ is the 
correlation factor which depends on the elastic-plastic properties
of the material. 
However, one should remember that at small length scales the 
mechanical characterization of materials indicates significant 
departures from the classical elastic–plastic behavior. 
Therefore, 
the characterization of SS by nanoindentation  can provide information
about its mechanical properties, if carefully performed \cite{LO200939,CHEN201783}. 

Concomitantly, Molecular Dynamics (MD) simulations have proven to
be a powerful tool to emulate experimental nanoindentation tests.
MD may provide atomistic insights to the mechanical response of 
indented samples and defect mechanisms
\cite{Voyiadjis2017, Yaghoobi2014, Mayo1990, Sato2020, Javier2021, KURPASKA2022110639,VARILLAS2017431}, 
information that cannot be clearly seen through Load - Displacement 
(L-D) curves. 
The major advantage of MD simulations is the ability to investigate 
the thermomechanical stability of dislocation nucleation and defects' 
evolution \cite{Voyiadjis2017, Yaghoobi2014, Mayo1990, Sato2020, Javier2021}. 
In addition, MD provides insights to the dislocations 
contribution on the relative increase or decrease of material
hardness \cite{Yaghoobi2014, Mayo1990}. 
Atomistic simulations can be further applied to study anisotropy in 
mechanical properties, providing a  predictive tool for 
experiments with prohibitive technical limits and costs. 
Thus, atomistic computational studies of  SS' nanomechanical 
response under external loads provides an insight into 
 fundamental defect mechanisms during testing that may explain the
 thermal dependence of plastic deformation, dislocation
nucleation rates, and strain-hardening rates. 

The present paper constitutes an attempt to fill in the gap related to multiscale computational modeling of mechanisms 
and plastic deformation of SS. 
We demonstrate a detailed experimental and computational study
to understand the nanoscale plastic deformation mechanisms and anisotropy effects in polycrystalline SS \cite{Zhang2019,Musial2022} and develop a multiscale description, using constitutive modeling.  
For the understanding of dislocation nucleation and 
evolution mechanism in SS during nanoindentation tests, we compare our MD results with the simpler, already studied, case of equiatomic FCC Ni-Fe solid solutions~\cite{KURPASKA2022110639} under similar thermomechanical 
conditions. Then, by visualizing and quantifying dislocation ensembles, we develop a continuum plasticity model for the defect evolution in SS.
Our manuscript is organized as follows: In Section 
\ref{sec:methods}, we describe the  experimental techniques and 
computational methodology for carrying out single load 
nanoindentation tests. 
By using MD simulations, we investigate
the mechanisms by which dislocations lines and dislocation loops
mediate plastic deformation at early stages of nanoindentation
in SS samples. 
In Section \ref{sec:results}, the hardness measurements and 
atomistic insights of indentation processes in crystalline 
stainless steel SS samples are presented, where an agreement 
between experimental measurements and numerical modelling is 
reported and discussed. Pop-in event identification is done by 
comparing to Hertz fitting curve in both methods. Qualitative good 
agreement is reached for the unloading process allowing us to apply 
Oliver-Pharr method. Finally, in Section \ref{sec:conclusion}, 
concluding remarks are summarized.
\section{Methods}
\label{sec:methods}
\subsection{Nanoindentation Experiments}

In this work, we study the mechanical properties of the low carbon 310S stainless steel with high chromium and nickel 
contents. This material is known for its resistance to high temperature corrosion \cite{LO200939,Zhang2019,Ustrzycka2012} due to the presence of these two elements. 

\begin{table}[b!]
\caption{Chemical composition of austenitic stainless steel 310S.}
\label{tab:Tab1}
\begin{tabular}{c llllllll}
\hline
Element   & Fe     & Cr    & Ni    & Mn 
& Si & C & P & S\\
\hline
weight \% & 53.096 & 24-26 & 19-22 & 2.0 
& 0.75 & 0.08 & 0.045 & 0.015\\
\hline
\end{tabular}
\end{table}

The chemical composition of our SS specimens
is presented in Table \ref{tab:Tab1}. The material was annealed
at $1100^{\rm o}$ C and air/water spray quenched. Studied specimens
in the shape of square plates with dimensions of about 
$1 \times 1$ cm were produced using electro-discharge machining 
(EDM) method. 
To reveal grain orientation, samples were submitted to the 
standard polishing procedure route (polishing with sandpaper 
from 320 till $4000\times$). The final step was done by 
electro-polishing using LectroPol 5 system. 
The described methodology allowed us to limit the hardening 
effect while effectively reducing surface roughness. 
Afterward, mechanical characterization by using nanoindentation
technique was performed using a NanoTest Vantage System provided
by Micro Materials Ltd. 
Measurements were done at room temperature using a 
Berkovich-shaped diamond indenter tip with the load control
method. Two basic parameters: hardness and reduced Young’s 
modulus with different indentation loads, were calculated. 
Before the indentation campaign started, the Diamond Area
Function (DAF) of the indenter, for each given load, was
determined by conducting a series of indentations on the fused
silica specimen (reference sample with well-known mechanical
properties). 
Indentations were performed using single force mode with loads
from 0.25 to 10 mN and were repeated at least 15 times at a
given load. The 60 sec thermal drift measurement time at the
end of each indentation cycle was recorded during the test. 
This was done to measure the thermal difference between the
sample and the indenter tip. 
The described methodology allowed us to investigate mechanical
properties as a function of the depth, hence taking into account
the response of one grain (for small loads) and the cumulative
effect of grains, grain boundaries, and precipitates (for
higher loads). 
Indentations were done with 20 $\mu$m distance between each
indents (in $X$ and $Y$ direction). This allowed us to avoid
interference of the indents or probing in an already deformed
(by previous indent) region which is particularly
important to prevent the influence of
indentation stress field. 
The experimental details of the nanoindentation campaign are
shown in Tab \ref{tab:Tab2}. 
Thus, the SS samples were characterized by calculating
the nanohardness (H) and reduced Young’s modulus (Er) values 
at different depths, following the well-known classical  
Oliver and Pharr approach \cite{oliver_pharr_1992}. 

\begin{table}[b!]
\caption{Experimental details of the single loads test (at room 
temperature).}
\label{tab:Tab2}
\begin{tabular}{c l}
\hline
Experiment type & Single load \\
\hline
Method & Load Controlled  \\
Load ramp control & Fixed time load \\
                  & and unload  \\
Max. load (mN) & 10  \\
Min. load (mN)      & 0.25  \\
Limit stop load (mN) & 0.1  \\
Indenter cont. vel. ($\mu$m/s) & 0.2\\
Load time (s) & 5 or 10 \\
Unload time (s) & 3 or 5 \\
Dwell period at max. load (s) & 1 or 2 \\
Dwell period for drift correction (s) & 60 \\
Number of columns & 1\\
Number of rows & 12 \\
Distance between indent ($\mu$m) & 20\\
\hline
\end{tabular}
\end{table}

\begin{figure}[t!]
   \centering
   \includegraphics[width=0.48\textwidth]{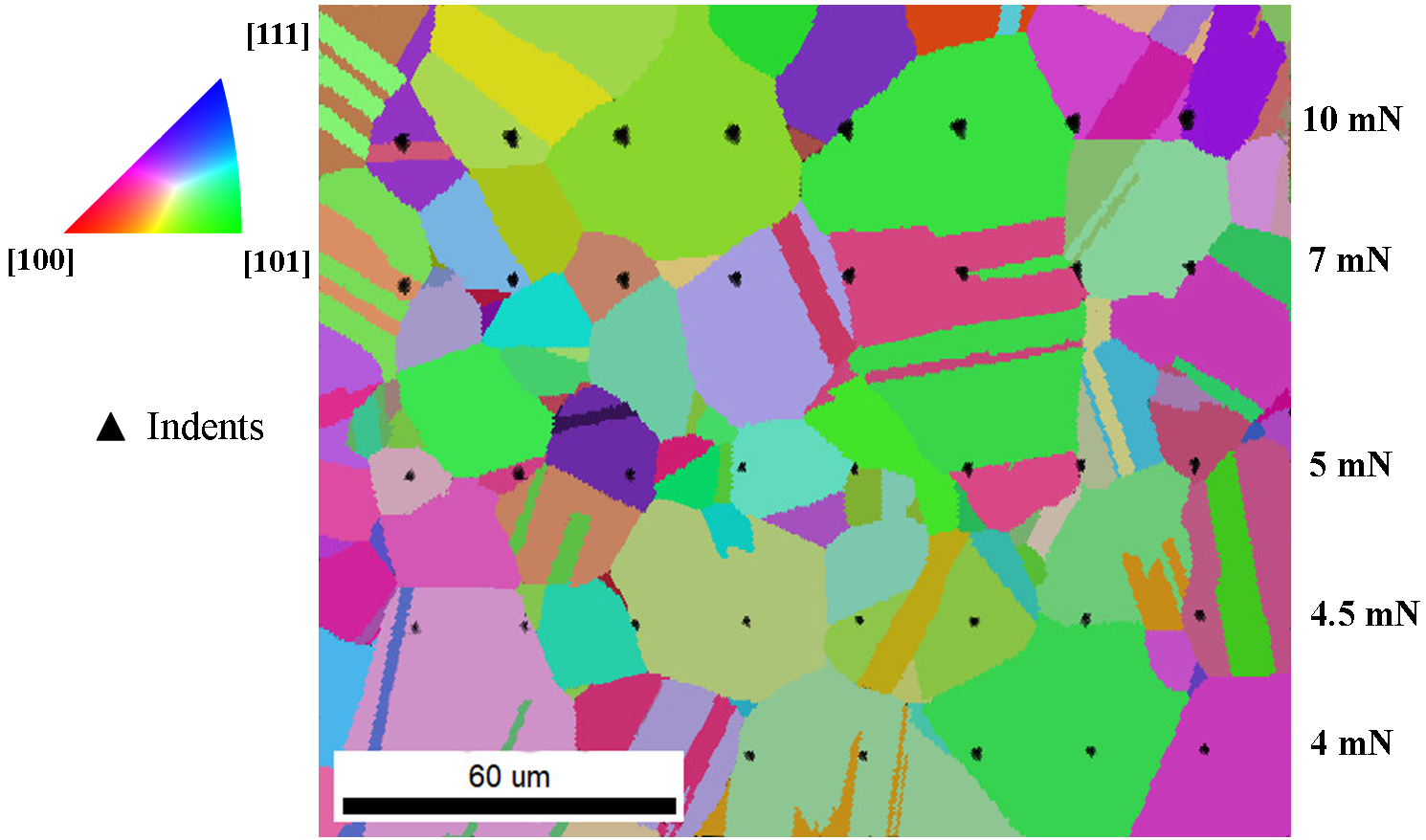}
   \caption{(Color on-line). Schematic of the standard configuration in the experiments for nanoidentation test. 
   An EBSD image was used to identify the orientation of grain 
   for the computational modeling.}
   \label{fig:fig0}
\end{figure}

After the mechanical test, detailed structural characterization
was done. The Electron Backscatter Diffraction (EBSD) analysis
of  the indented sample was conducted using ThermoFisher 
Scientific Helios 5 UX Scanning Electron  Microscope (SEM) 
equipped with an EDAX Velocity Pro EBSD system. 
The mapping of the specimens was done using 20 keV electron 
beam of 6.4 nA probe current. 
The grain reconstruction in the collected EBSD maps has been performed 
in EDAX OIM Analysis 8 software by an algorithm that groups sets 
of connected and similarly oriented points into grains 
if they are within specified Grain Tolerance Angle (equal to 5 deg. 
in the studied case) of a given point. 
The Minimum Grain Size, which is the number of points on the 
measurement grid required whether a given group of points should 
be considered a grain, has been set to 2. 
A crystallographic orientation expressed in $(hkl)[uvw]$ can be 
assigned to each reconstructed grain. 
This information for selected grains has been used as input data 
for the computational modeling, with approximately identical 
orientations. The average grain size of the as-received SS
was around 20 $\mu$m, as depicted in Fig \ref{fig:fig0}.  

\subsection{Atomistic modeling and simulations}

Atomistic computational modeling is based on the Molecular Dynamics 
(MD) simulations by the Large-scale Atomic Molecular
Massively Parallel Simulator (LAMMPS) \cite{THOMPSON2022108171} 
with interatomic potentials based on EAM \cite{Bonny_2013}
to describe the atom-to-atom interactions in SS. 
In order to model FCC Fe-Ni-Cr samples, we first create a pure FCC 
Fe sample with a lattice constant of 0.3562 nm followed by randomly 
replacing Fe atoms by Ni and Cr atoms at [100], [110], and [111] 
crystal orientations. 
However, the potential energy of structure is minimized, with the 
idea the real value is achieved in the process. 
The lattice constant is slightly changed during the sample 
equilibration compared to the originally adopted. 
The obtained sample with an atomic distribution of: 54.7\% Fe; 
18.8\% Ni; and 
26.5 \% Cr is then prepared by a series of Monte Carlo simulations 
to search for each possible
metastable configuration at room temperature. 
Then, we applied a process of energy optimization and  
equilibration for 100 ps with a Langevin thermostat at 300 K
and a time constant of 100 fs \cite{Mayo1990}. 
This is done until 
the system reaches a homogeneous sample 
temperature and pressure profile \cite{Mayo1990} with a density of 
$8.0$ g/cm3, the numerical parameters that defined our 
numerical cells are presented in Tab \ref{tab:Tab3}. 
For the equiatomic Ni-Fe concentrated solid solution Alloy, we utilized 
a potential interatomic reported by Choi et al. \cite{Choietal} which 
are based on the second Nearest Neighbor Modified Embedded Atom Method (2NN-MEAM). 
The numerical cell defined as: (33.08x, 35.95y, 30.97z) 
and with 1 582 400 Ni and 1 582 400 Fe atoms, as reported in our previous 
work \cite{KURPASKA2022110639}.

\begin{table}[b!]
\caption{Size and atomic distribution of the numerical samples used to perform MD simulations.}
\label{tab:Tab3}
\begin{tabular}{c ll}
\hline
Orientation   & Size(dx,dy,dz) [nm] & Atoms   \\
\hline
$[001]$ & (44.68,43.97,50.09) & 8 610 000 \\
$[110]$ & (44.58,43.78,52.62) & 8 985 600 \\
$[111]$ & (44.73,43.78,52.67) & 9 027 000 \\
\hline
\end{tabular}
\end{table}

\begin{figure}[t!]
   \centering
   \includegraphics[width=0.45\textwidth]{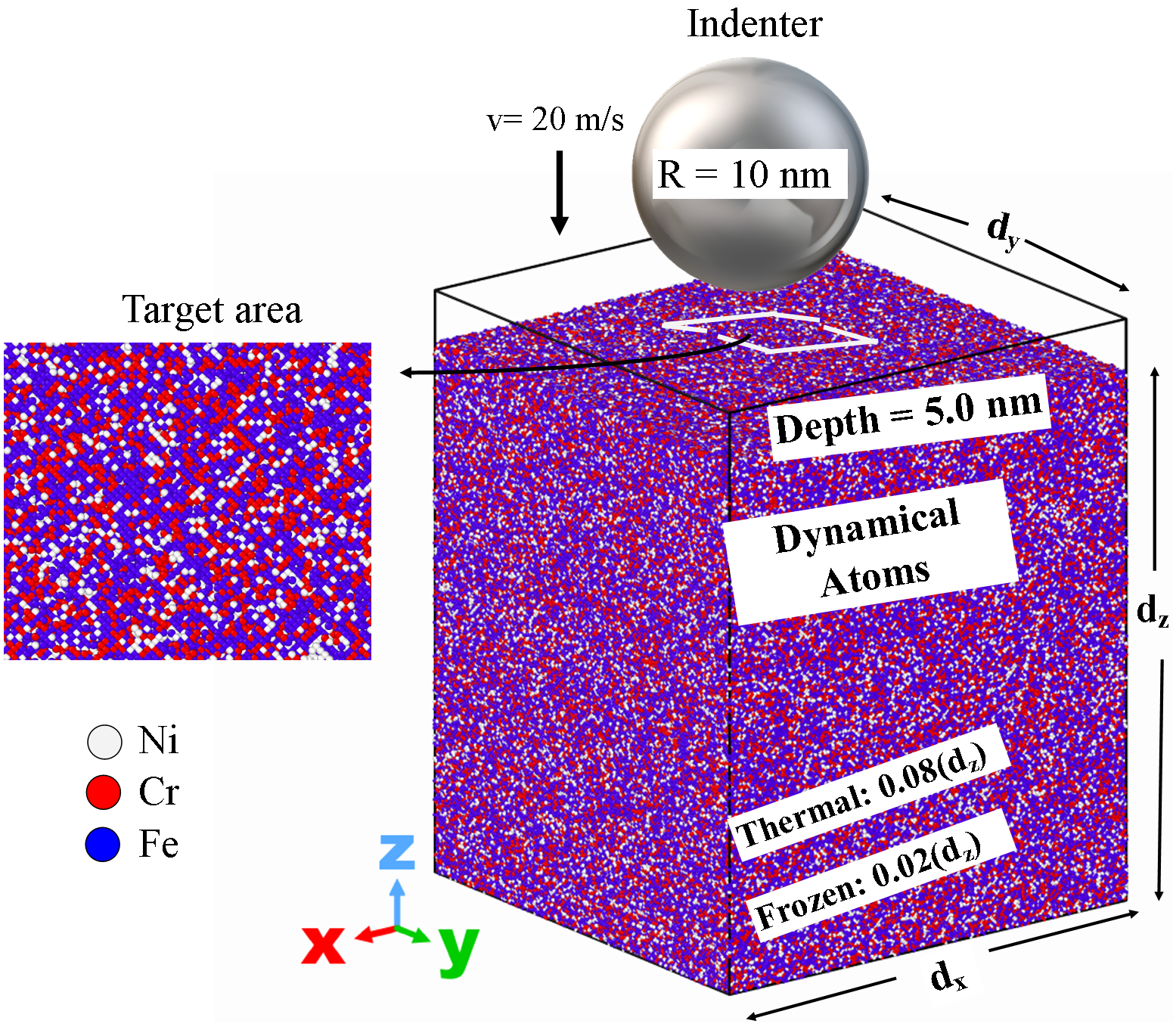}
   \caption{(Color on-line). Schematic of the standard configuration in our MD 
   simulations of nanoindentation test. 
   The prepared 310S specimen is divided into three regions to consider boundary
   conditions and a non-atomic repulsive spherical indenter is used. 
   Configuration of the target area shows the random 
   atomic distribution of the sample.}
   \label{fig:fig1}
\end{figure}

Under the spatial (few nanometers) and time (picoseconds) scales
of the MD simulations, the initial processes of the plastic
deformation of the materials can be studied at the atomic level
by nanoindentation testing \cite{Nix1998}. 
In addition, the computational modeling is an approximation to
the  roundness of the Berkovich tip into the consider errors due
to the use of a spherical indenter in the MD simulations, which
is  limited to only few nm depths \cite{PhysRevLett.95.045501}. 
Furthermore, the 20 m/s indentation velocity considered in our
work is smaller than the sound’s speed in solids where 
our computational results can accurately capture the elastic 
Hertzian regime and provide information of early
dislocation nucleation, similar to those obtained in 
experiments, along with a better understanding of the 
elastic-plastic deformation transition of the material.  
In Fig \ref{fig:fig1}, we present the initial frame of the
nanoindentation  simulation which is defined into three  
sections in the $z$ direction. Thus 1) the lowest bottom 
layers are kept frozen ($\sim$0.02$\times$dz) to assure stability
of the atoms when nanoindentation is performed;  2) a 
thermostatic region ($\sim$0.08$\times$dz) above the frozen one
is set to dissipate the generated heat during nanoindentation;
and 3) the rest of 
the layers are defined as the dynamical atoms section,  where
the interaction with the indenter tip modifies the
surface structure of the samples.  Finally, a 5 nm vacuum section
is added at the top of the sample \cite{Javier2021}.

In our work, we use a NVE statistical thermodynamic ensemble
to carry out the indentation test, where the velocity Verlet
algorithm is implemented in LAMMPS with periodic boundary 
conditions set on the $x$ and $y$ axes to simulate an infinite
surface. 
A non-atomic repulsive imaginary (RI) rigid sphere defines our
indenter tip as: $F(t) =K\left(r(t)-R\right)^2$ where the 
constant force is K= 37.8 MPa, and the trajectory of the center
of the indenter tip is defined as $r(t) = (x_0, y_0, 
(z_0 \pm vt))$, with radius $R$= 10 nm, surface contact point
as $x_0$ and $y_0$, and the initial gap $z_0 = 0.5$ nm
between the surface and the intender tip to avoid effects
of initial tip-surface interaction. 
The indenter tip’s speed $v = 20$ m/s is chosen as positive
for loading, and as negative for unloading processes. 
Each calculation was performed for 125 ps with a time step
of $\Delta t = 0.5$ fs. for a maximum indentation depth of
5.0 nm to avoid the influence of boundary layers of the
material. 
We consider the random atomic distribution of the elements 
on the material surface by performing $N_{\rm MD}=$ 10 
simulations at different indenter tip's positions 
into 10nm$\times$10nm target area, as depicted by 
a white square in Fig \ref{fig:fig1}. 
Thus, the load on the indenter $P$ is computed by the forces
acting on the indenter in the $z$-axis direction and the
depth $h$ is calculated as the displacement of the indenter
tip relative to the initial surface of the material sample.

\subsection{Measures of Hardness and Elastic Moduli}
Once the experimental and computational nanoindentation test 
is done, the unloading curve of the load displacement recording 
is analsyzed to apply the Oliver-Pharr method as follows:
\begin{equation}
    P=P_0 \left(h-h_f \right)^m
    \label{equation1}
\end{equation}
with $P$ is the indentation load; $h$ is the indentation depth
and $h_f$ is the residual depth after the whole indentation process;
and $P_0$ and $m$ are fitting parameters. Thus, the nanoindentation 
hardness can be computed as: 
$H = P_{\rm max}/A_c$ where $P_{\rm max}$ 
is the maximum indentation load at the maximum indentation
depth, 
$A_c = \pi(2R - h_c)$ where $h_c$ is the projected contact area
with
$R$ as the indenter tip radius and $h_c = h_{\rm max} - \epsilon 
P_{\rm max}/S$. 
Here $\epsilon = 0.75$ is a factor related to the spherical
indenter 
shape for the MD simulations, and unloading stiffness $S= dP/dh$. 
Thus, the Oliver-Pharr method can be applied to compute the Young's modules 
of the material at different 
indentation depths as \cite{oliver_pharr_1992,OliverPhar2}.

The Young's module $E_{\rm Y}$ is computed as:
\begin{equation}
    \frac{1-\nu^2}{E_{\rm Y}} = \frac{1}{E_{\rm r}} - 
    \frac{1-\nu_{\rm i}^2}{E_{\rm i}},
\end{equation}
where $\nu$ and $\nu_{\rm i}$ are the Poisson's ratio of the 
310S sample and indenter, respectively. 
$E_{\rm i}$ is the Young's modulus of the spherical indenter 
that is considered to be infinitely large, 
and the effective elastic modulus $E_{\rm r} = \sqrt{\pi/A_c}S/2\beta$ 
with $\beta = 1$ and $1.034$ for a spherical and Berkovich 
indenter shape, respectively \cite{OliverPhar2}.

\subsection{Atomic shear strain mapping}
\label{sec:appendixB}
For the shear dependence of nanoindentation, atomic strains are computed through the 
distance difference, $\textbf{d}^{ \beta}$, between the the 
$m$-th nearest neighbors of the $n$-th atom of the 
pristine and indented samples. 
Followed by defining the Lagrangian strain matrix 
of the $n$-th atom as
\cite{2007MJ200769}:
\begin{eqnarray}
    \boldsymbol{\eta}_n = & 1/2 \left(\boldsymbol{J}_{n} \boldsymbol{J}_{n}^T-I\right), \\ 
      & {\rm with} \nonumber \\
    \boldsymbol{J}_{n} = & \left( 
    \sum_{m}\boldsymbol{d}_{m}^{0T}\boldsymbol{d}_{m}^{0} 
    \right)^{-1} 
    \left(\sum_{m}\boldsymbol{d}_{m}^{0T}
    \boldsymbol{d}_{m} \right).
\end{eqnarray}
Thus, the shear invariant of the $n$-th atom is computed as:
\begin{equation}
    \eta_n =   \sqrt{\frac{\zeta_{ij}\zeta_{ij}}{2}}, \quad 
    {\rm with} \quad
    \zeta_{ij} =  \eta_{ij}-\eta_{kk} \delta_{ij}. 
\end{equation}
This approach is implemented in OVITO \cite{ovito}.

\subsection{Multiscale dislocations evolution}
\label{subsec:m-disloc}

In order to analyze the influence of the crystal orientation on
the dislocation nucleation and evolution of the sample, we visualize and quantify
 different types of dislocations nucleated at different 
indentation depths by using the OVITO \cite{ovito} software. This was done through the use of the Dislocation Extraction Algorithm (DXA)~\cite{Stukowski_2012}; that extracts dislocation structure and content from atomistic microstructures. 
Thus, we categorized the dislocations into several dislocation
types according to their Burgers vectors as: ½<110> (Perfect),
1/6<112> (Shockley), 1/6<110> (Stair-rod), 1/3<100> (Hirth),
1/3<111> (Frank) noticing that the nucleation of partial 
1/6<112> Shockley dislocations is dominant in the loading 
process regardless of the crystal orientation due to the material's
FCC structure. 
Thus, we compute the dislocation  density, $\rho$, as a function
of the depth as

\begin{equation}
    \rho = \frac{N_D l_D}{V_D},
\end{equation}

where $N_D$ is the number of dislocation lines and loops measured during 
nanoindentation test; $l_D$ is the dislocation length of each type, 
and $V_D = 2\pi/3(R_{\rm pl}^3-h^3)$ is the volume of the plastic 
deformation region by using the approximation of a spherical 
plastic zone; where $R_{\rm pl}$ is the largest distance of 
a dislocation measured from the indentation displacement, 
considering a hemispherical geometry. 
In order to obtain more information about the nanomechanical response of
the material during loading event, we calculate the indentation
stress and strain by considering the contact radius between the
sample and the tip by using the geometrical relationship $A_C =
\sqrt{R_i^2-(R_i-h)^2 }$. 
Thus, the nanoindentation stress and strain are calculated using
the following equations \cite{PATHAK20151}:

\begin{equation}
\sigma_{\rm IT} = \frac{P}{\pi a^2} \quad {\rm and} 
\quad \epsilon = \frac{4h}{3\pi a}
\end{equation}

where $P$ is load, $h$ is indenter displacement. 
The first expression for the nanoindentation strain is considered
as a physical strain where the nanomechanical response of the 
material is provided by the indentation depth ($h$) and the 
contact radius of the indenter tip, $a$. 

The following constitutive laws of dislocation mechanics are used
to describe the evolution of dislocation density during plastic
deformation in continuum material point \cite{Kock2003, Tsuchida2001}
\begin{equation}
    \frac{d \rho}{d \gamma}={\frac{d \rho}{d \gamma}}\Big|_{+}
    +{\frac{d \rho}{d \gamma}}\Big|_{-} 
      \label{Eq:Eq4}
\end{equation}
with
\begin{equation}
     \frac{d \rho}{d \gamma}\mid_{+}=  \frac{1}{\lambda b}  
      \label{Eq:Eq5}
\quad {\rm and} \quad
\frac{d \rho}{d \gamma}\mid_{-}=  -k_a \rho 
\end{equation}
where $b$ denotes length of the Burgers vector, $\lambda$ is the
mean free path of dislocation and $k_a$ is the dislocation annihilation constant \cite{Bouaziz2012}. The evolution of the total dislocation density $\rho$ during plastic deformation is decomposed into the component of dislocation multiplication denoted by $(+)$ associated with the production of new dislocations and the annihilation component denoted by $(-)$.
The relation between the effective plastic strain rate 
$\Dot{\epsilon}$ and the the plastic shear rate $\Dot{\gamma}$ 
is expressed by a mean orientation factor $M$ as 
$\Dot{\gamma}=\Dot{\epsilon}M$.
As a result the relationship between the evolution of $\rho$
and the plastic strain is obtained in the following form
\begin{equation}
\dot{\rho}= \left ( \frac{1}{bd}+\frac{k_1}{b} \sqrt{\rho} -k_a \rho \right)  \dot{\gamma}
  \label{Eq:Eq9}
\end{equation}
where $d$ is the average grain size and $k_1$ is a constant.
\section{Results}
\label{sec:results}
\subsection{Nanoindentation experiments and molecular simulations}

\begin{figure}[b]
   \centering
   \includegraphics[width=0.40\textwidth]{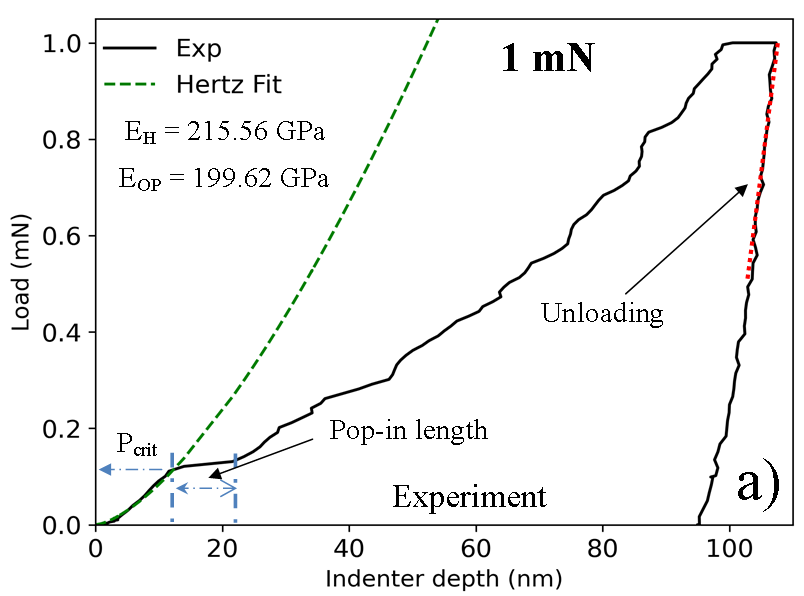}
   \includegraphics[width=0.40\textwidth]{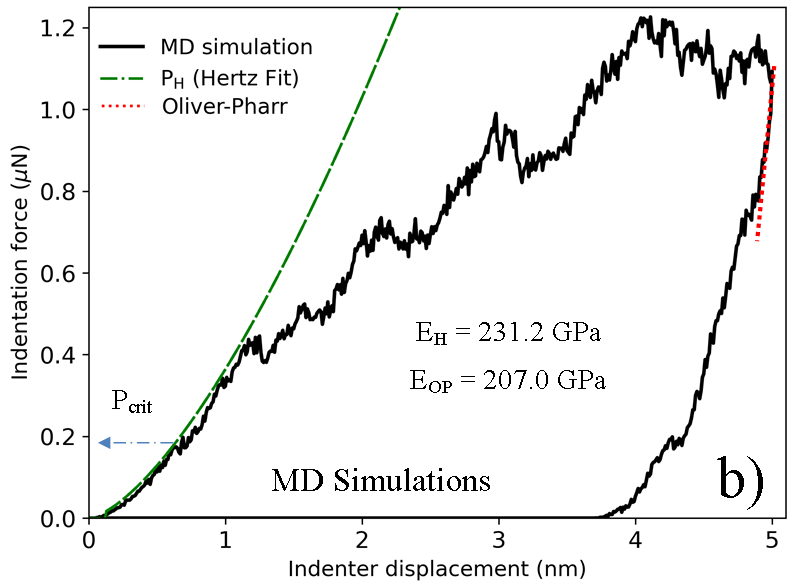}
   \caption{(Color on-line). LD curve obtained experimentally in a) and from the
   numerical modelling on the [110] crystal orientation in b). Pop-in event 
   identification is done by comparing to Hertz fitting curve in both methods. 
   Qualitative good agreement is reached for the unloading process allowing us
   to apply Oliver-Pharr method.}
   \label{fig:fig2}
\end{figure}

In order to numerically model the nanoindentation test with a single
force mode, we need to follow the recorded LD curve that is obtained
experimentally as shown in Fig \ref{fig:fig2}a). 
Here we present an initial loading-unloading curve recorded at 1mN reaching a depth
of 100 nm. 
The test was done at  a [110] grain, according to the EBSD image shown in
Fig \ref{fig:fig0}. 
At the beginning of the loading process, the pop-in
event~\cite{xu,Papanikolaou2017} is identified by fitting this part of the curve to the Hertz curve defined as: 
$P_H=4/3 E_{\rm H} R^{1/2} h^{3/2}$ we calculate the effective elastic modulus
$E_{\rm H}$ of the studied system. 
This defines the 
critical load (and length) of the pop-in which is necessary to trigger
elastic-plastic transition \cite{PhysRevLett.106.165502}.
In Fig \ref{fig:fig2}b), we show results for the numerical 
modeling of nanoindentation test performed on the [110] 
crystal orientation. One can observe that the pop-in event guides
the elastic-plastic transition which is observed as a deviation of the force
respect to the Hertz fitting curve, (without pop-in length). 
Calculated experimental elastic modulus value for this selected LD curve is 
$E_{\rm H}$ = 215.56 GPa, while the MD 
simulation results with $E_{\rm H}$ = 231.2 GPa.
In addition to that, both methods allowed us to observe several pop-in events during the nanoindentation test. Also, one can see that the unloading process recorded experimentally and simulated is in qualitative good agreement, 
as shown in Fig \ref{fig:fig2}. 
Thus, the Oliver-Pharr method is applied to compute the hardness 
and Young's modulus of the material by applying a fitting curve 
to the unloading curve in both methods as shown 
in Sec. 2.2.

\begin{figure}[b!]
   \centering
   \includegraphics[width=0.43\textwidth]{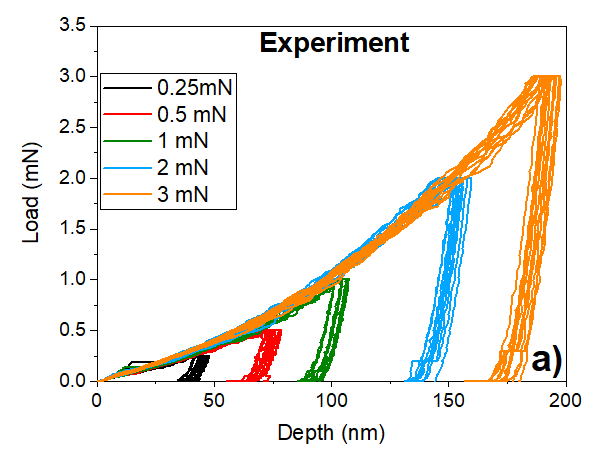}
   \includegraphics[width=0.40\textwidth]{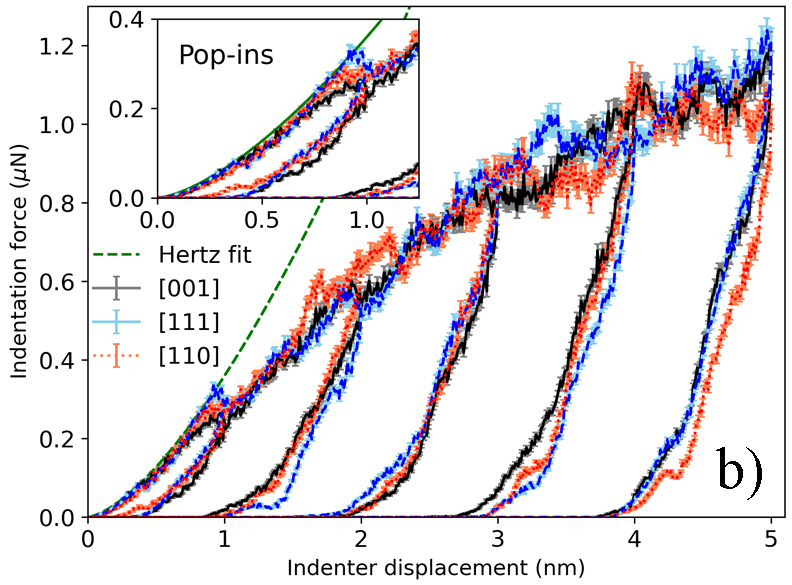}
   \caption{(Color on-line). LD curves from MD simulations in a) for different crystal orientations. Experimental data obtained by single load nanoindentation test at different initial load in b).}
   \label{fig:fig4}
\end{figure}

In Fig \ref{fig:fig4}a), we report the recorded LD curves for 
different initial load values to investigate the hardness of the 
sample material as a function of its depth for several indentation tests.
This was done to collect
more statistical data and take into account
morphology of the surface, crystal orientation and impact of grain boundaries, as
these events must be taken into account for 
this kind of mechanical tests. 
As shown by the experiments, at a load range of 0.25 to 3 mN where 
plastic deformation can be at some extend compared to MD simulations, albeit that
the spherical indenter tip is moved at a fixed speed  during the computational
nanoindentation test. 
In order to emulate the experiments, we performed MD simulations reaching different
depths in the material so that the Oliver-Pharr  method can be applied to each
indentation depth. 

In Fig \ref{fig:fig4}b), we present the results from 10 MD simulations 
for each [001], [110], and [111] crystal orientations taking into 
account the random atomic distribution on the material surface. 
Thus, the average of the results is reported as: 
$P=1/n (\sum_nP_n )$ where $N_{\rm MD}=10$ is the number of the MD
simulations and the error bars correspond to the maximum and
minimum values from the simulations. 
We noticed that the maximum pop-in load has been recorded for the [111] 
orientation, followed by [110], while the lowest value defines the [001]. 
This is representative for FCC samples (shown in the inset figure).
During the loading process several pop-ins are observed in the MD 
simulations due to the distortion of the lattice constant of the 
material. 
Experimental data show also these effects where several pop-ins 
are observed at depths smaller than $30$nm. 
Recorded data suggest that the correlations between the crystal 
orientation and the pop-in magnitude exists. 
This will be further analyzed in the next section of the article. 

\begin{figure}[b!]
   \centering
   \includegraphics[width=0.48\textwidth]{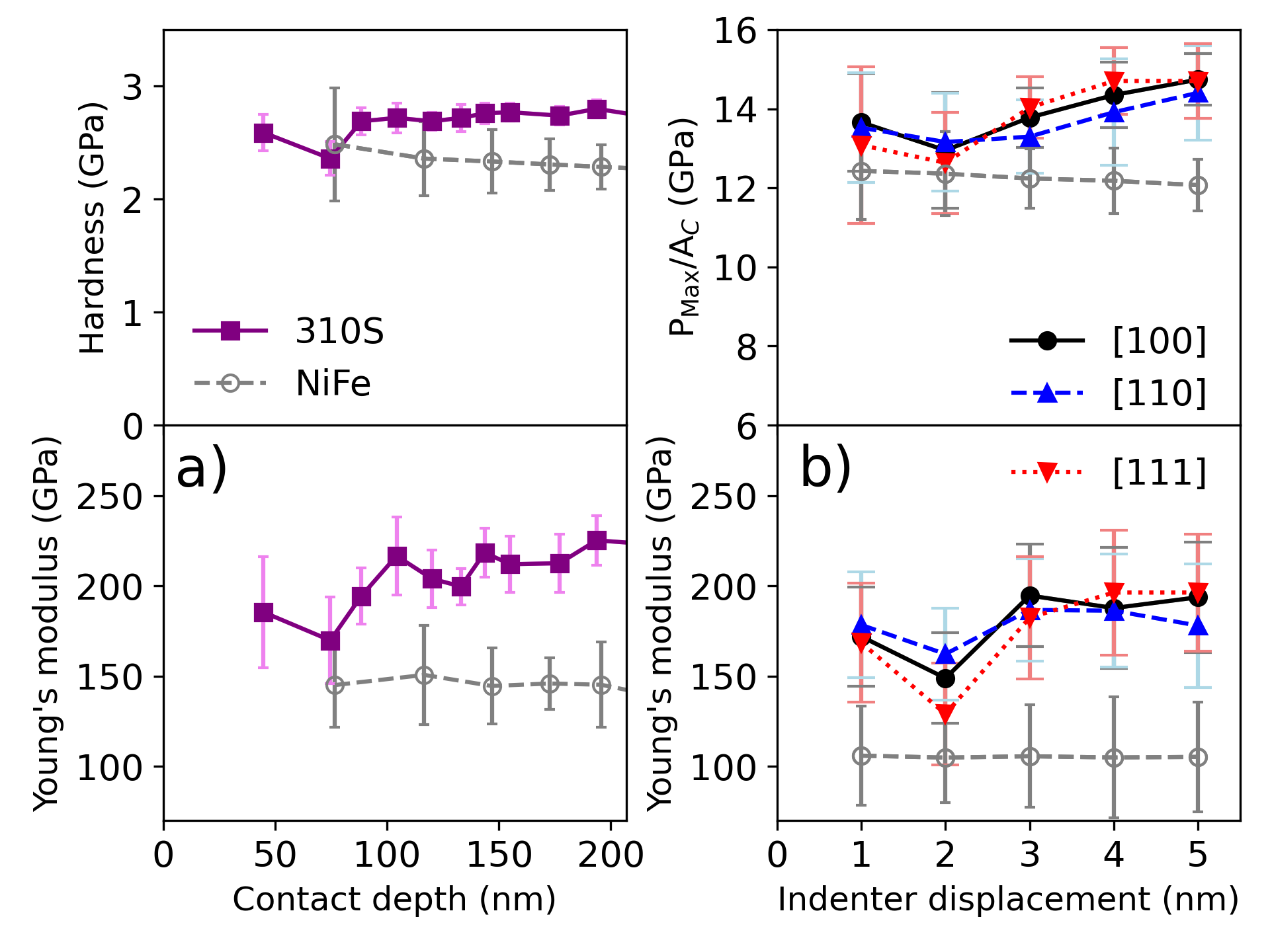}
   \caption{(Color on-line). Hardness vs depth from experiments 
   in a) and from MD simulation in b) for different crystal 
   orientation, a qualitative agreement is reached by both methods.
   A comparison to equatomic NiFe SP-CSA is included to show the
   effect of chemical complexity on the mechanical properties
   of the materials
   \cite{KURPASKA2022110639}.}
   \label{fig:fig5}
\end{figure}

In Fig \ref{fig:fig5}, we present results of the hardness and 
Young's modulus vs indentation depth recorded experimentally, in Fig \ref{fig:fig5}a),
and from the numerical modeling in b). In the numerical part we consider different
crystal orientations. 
The error bars are associated to the maximum and minimum value obtained 
from 10 MD simulations. 
We noticed a qualitatively good agreement between both methods 
where recorded pop-in event during the loading process can be associated to the 
nucleation of geometrically necessary dislocation (GND), 
as analyzed in the following sections. 
Presented results were obtained by applying the Oliver-Pharr method 
to the unloading curve at different initial loads 
(Fig \ref{fig:fig2}a), while the MD simulations provide information
of the unloading process at each indentation depth (with single nm precision)
where the calculations of the hardness and Young's modulus were done. 
Same as the experimental data, MD simulations present higher error
bars for small depths. Also, one can see that the calculated hardness and young
modulus values are within the error bars, regardless of the studied crystal 
orientation. 
This is related to the sample roughness (in the case of experimental tests) and
probably to the orientation of the material (in the case of simulations) where
the nanoindentation tests were carried out.
Hardness data from experiments are reported in the supplementary material. 
In order to report the effects on chemical complexity of 
the material on the mechanical properties, we compare results 
to those reported for an equiatomic single-phase concentrated 
solid solution alloy (SP-CSA) defined as NiFe \cite{KURPASKA2022110639} 
for experimental and computational modeling. 
We noticed that the hardness of the binary alloy smoothly 
decreases as a function of the depth for both methods. 
Reduced Young's modulus is constant regardless the depth 
of the tip. 
This comparison shows the effect of lattice mismatch 
and Cr concentration in 2 to 3 elements mixed materials.

\begin{figure}[t]
   \centering
   \includegraphics[width=0.48\textwidth]{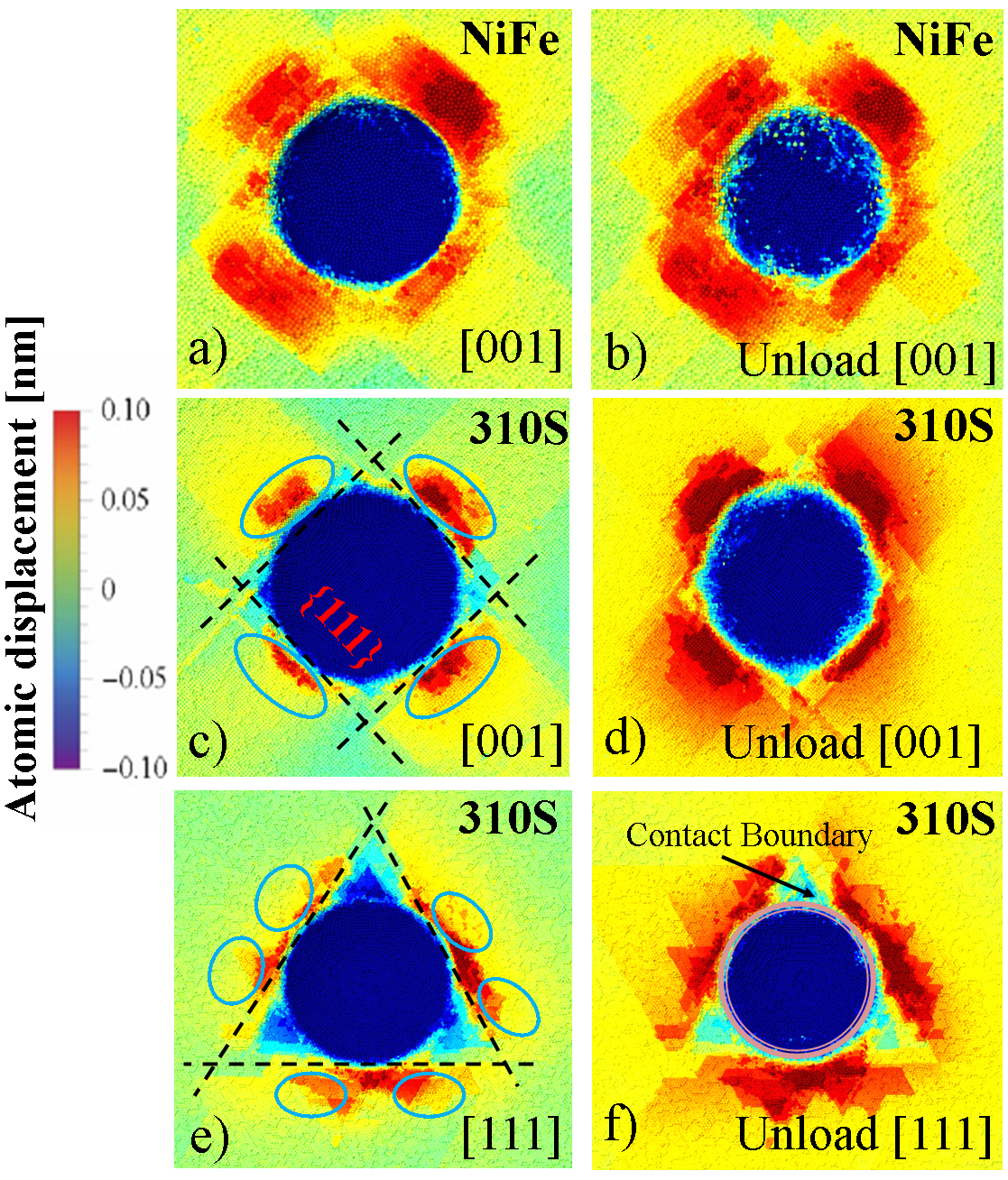}
   \caption{(Color on-line). Simulated contour plot of pileups 
   formed at the maximum indentation depth for [001] NiFe in a-b) 
   \cite{KURPASKA2022110639}, and SS sample for [001] in c-d) and 
   [111] in e-f) showing the surface topography 
   after indentation and simultaneously exposes plastic 
   deformation of the surface.
   We show the slip system family and the out-of-plane 
   displacements are highlighted by blue circles.}
   \label{fig:fig3}
\end{figure}

The simulated nanoindentation tests were carried out by considering different 
crystal orientations to investigate their effects
on the mechanical properties of the material and account for surface morphology. 
In Fig \ref{fig:fig3}, we present the atomic displacement at 
the maximum indentation depth from the MD simulations calculated
for [001] NiFe in a-b), and for SS sample for the crystal
orientations of [001] in c-d) and [111] in e-f). 
We identified that the \{111\} slip system family activated 
during the loading process which is associated to the slip 
traces 
on the [110] orientation.
The out-of-plane displacements are highlighted by blue
circles presenting the typical morphology for FCC materials
\cite{Sato2020} where the profile around the indent on the 
[001] orientation shows a fourfold symmetry, while the 
[111] orientation is defined by the sixfold symmetry. 
Formed pile-ups following \{111\} planes are observed at the
maximum indentation depth. 
It is worth pointing out, that the pile-ups formation results
primarily due to the crystalline nature of the material and
depends strongly on the hardening of the material during
deformation.

\begin{figure}[bth]
   \centering
   \includegraphics[width=0.35\textwidth]{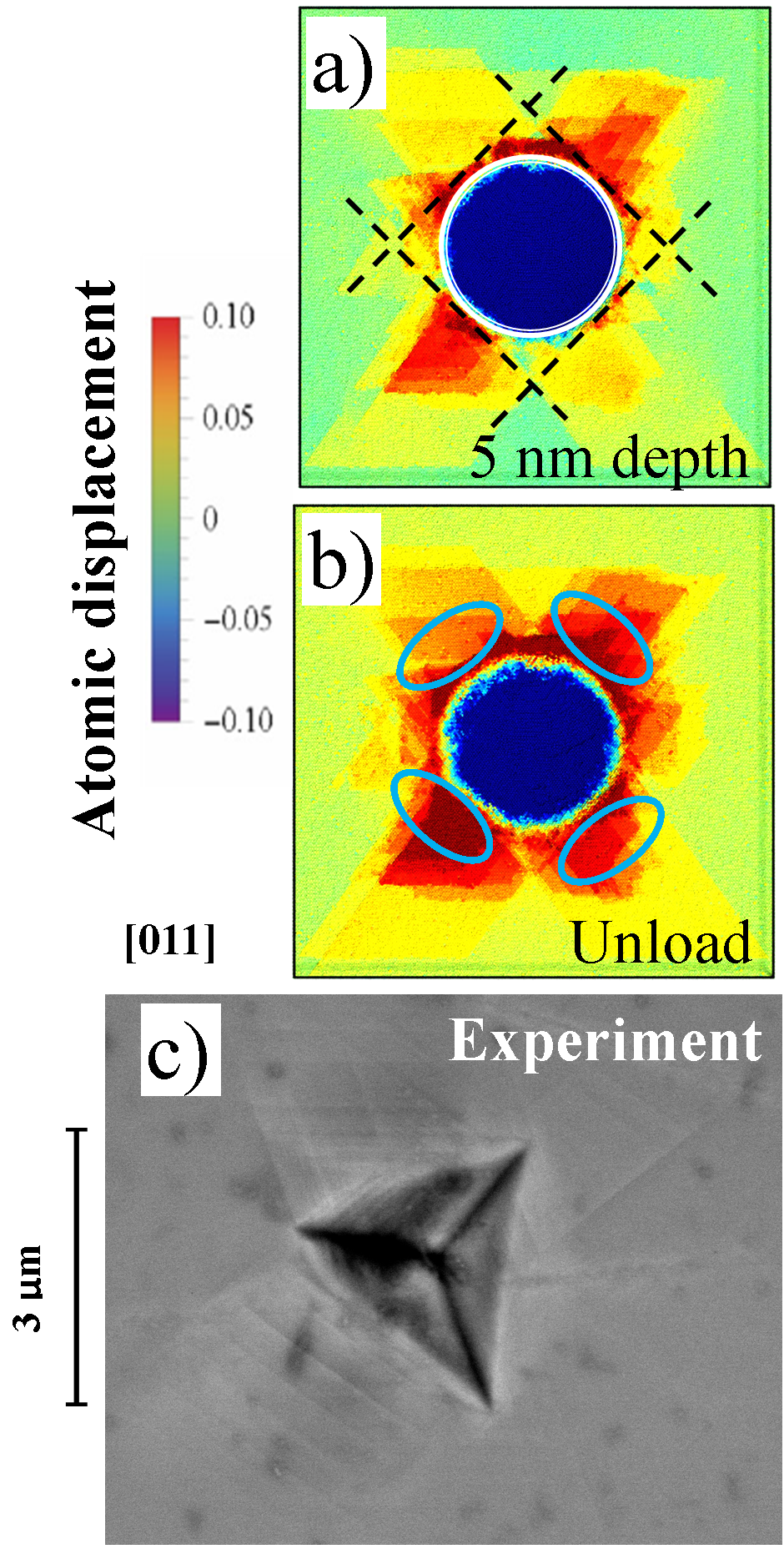}
   \caption{(Color on-line).
   Atomic displacement at the maximum indentation depth in a) 
   and after unloading process in b) for [110] crystal orientation.
   SEM image of slip traces for an indent on a [110] grain 
   boundary in c).
   Qualitative good agreement is reached by both methods 
   showing the same crystallographic slip directions on 
   $\{112\}$ and $\{110\}$ symmetry family planes.}
   \label{fig:fig3-b}
\end{figure}

In order to qualitatively compare MD simulations results with 
experimental data in Fig. \ref{fig:fig3-b} we present 
the atomic displacement for the [110] orientation at the 
maximum indentation depth in a) and after nanoindentation test 
in b); SEM image of slip traces formation for an indent 
on a [110] grain in c). 
These images show a good agreement where the two-fold symmetry 
and the attainment of crystallographic slip directions are observed
by both methods. 
Experimental data for this indent at 225 GPa agrees well with MD simulations 
report in Fig. \ref{fig:fig4}
\subsection{Dislocation mechanisms and multiscale modeling}
\label{subsec:disloc}

\begin{figure}[t!]
   \centering
   \includegraphics[width=0.45\textwidth]{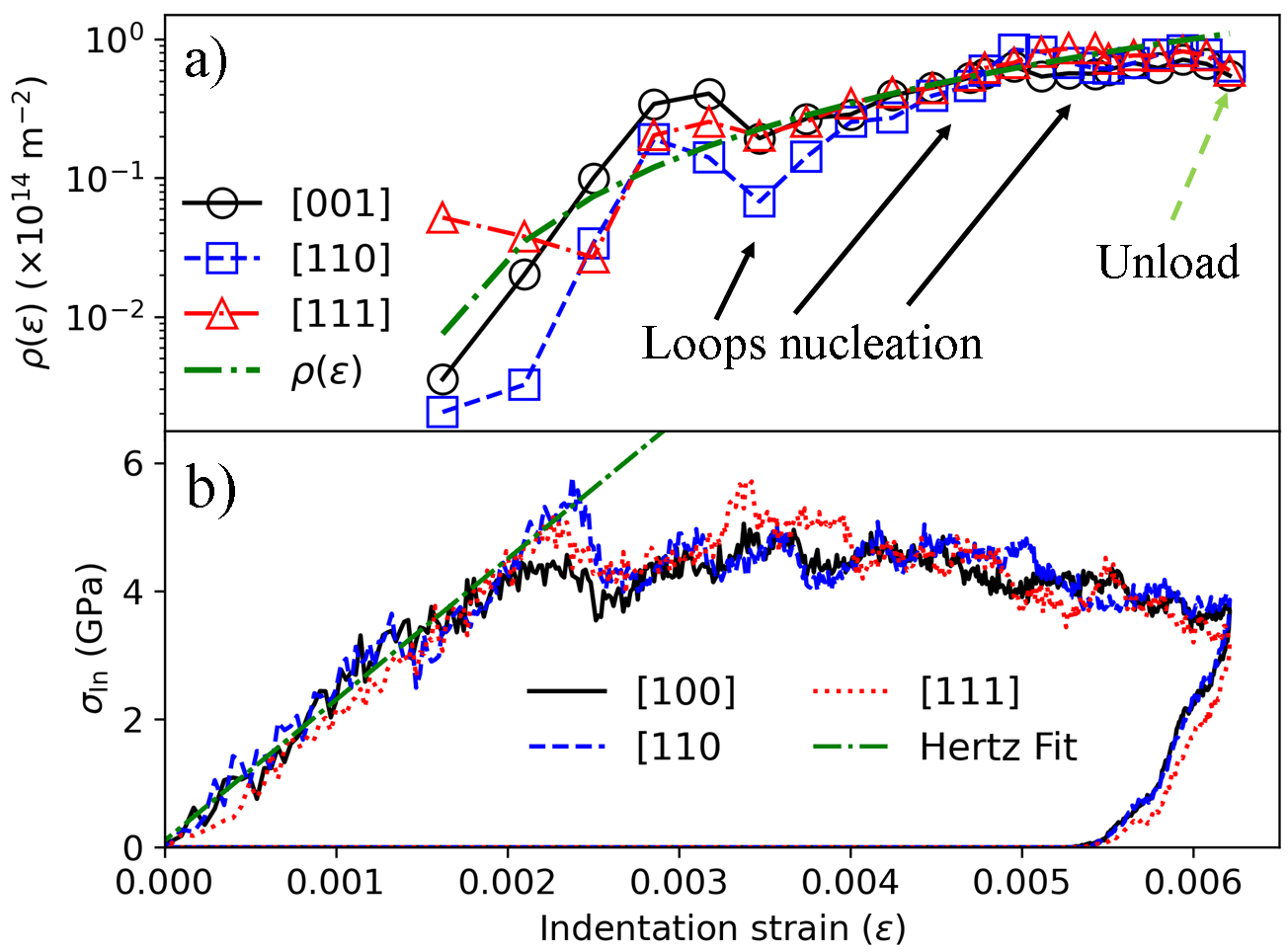}
   \caption{(Color on-line).Dislocation density at different crystal orientation 
   as a function of the nanoindentation depth in a) and indentation strain in b). Kocks-Mecking model, $\rho(\epsilon)$, is added in 
   a) by using Eq (\ref{Eq:Eq9}).}
   \label{fig:fig7}
\end{figure}

In Fig \ref{fig:fig7} we show results of the average dislocation 
density in a) and indentation stress in b) calculated for different 
crystal orientations as a function of the indentation strain. 
The interaction of symmetrical Shockley dislocations are observed
to be responsible for prismatic dislocation loops (PDL) nucleation, 
as identified in Fig \ref{fig:fig7}a) for different crystal
orientations. 
In Fig \ref{fig:fig7}b), we show the dislocation density as a function of the indentation strain for different crystal orientations 
where the elastic process is fitted to Hertz equation identifying the first pop-in event.

A comparison of the evolution of the dislocation density for 
different crystal orientations as a function of the indentation
strain obtained from MD simulations with theoretical curve (Eq
\ref{Eq:Eq9}) is shown in Fig \ref{fig:fig7}. 
The agreement can be regarded as satisfactory. 
This also shows that the adopted constitutive assumptions 
(Eqs. \ref{Eq:Eq4} and \ref{Eq:Eq5}) are sufficient for the
description of kinetics of dislocation density.
This equation is assumed to principally reflect the dislocation
mechanisms governing inelastic deformation in the indented material. 

\begin{figure*}[t!]
   \centering
   \includegraphics[width=0.85\textwidth]{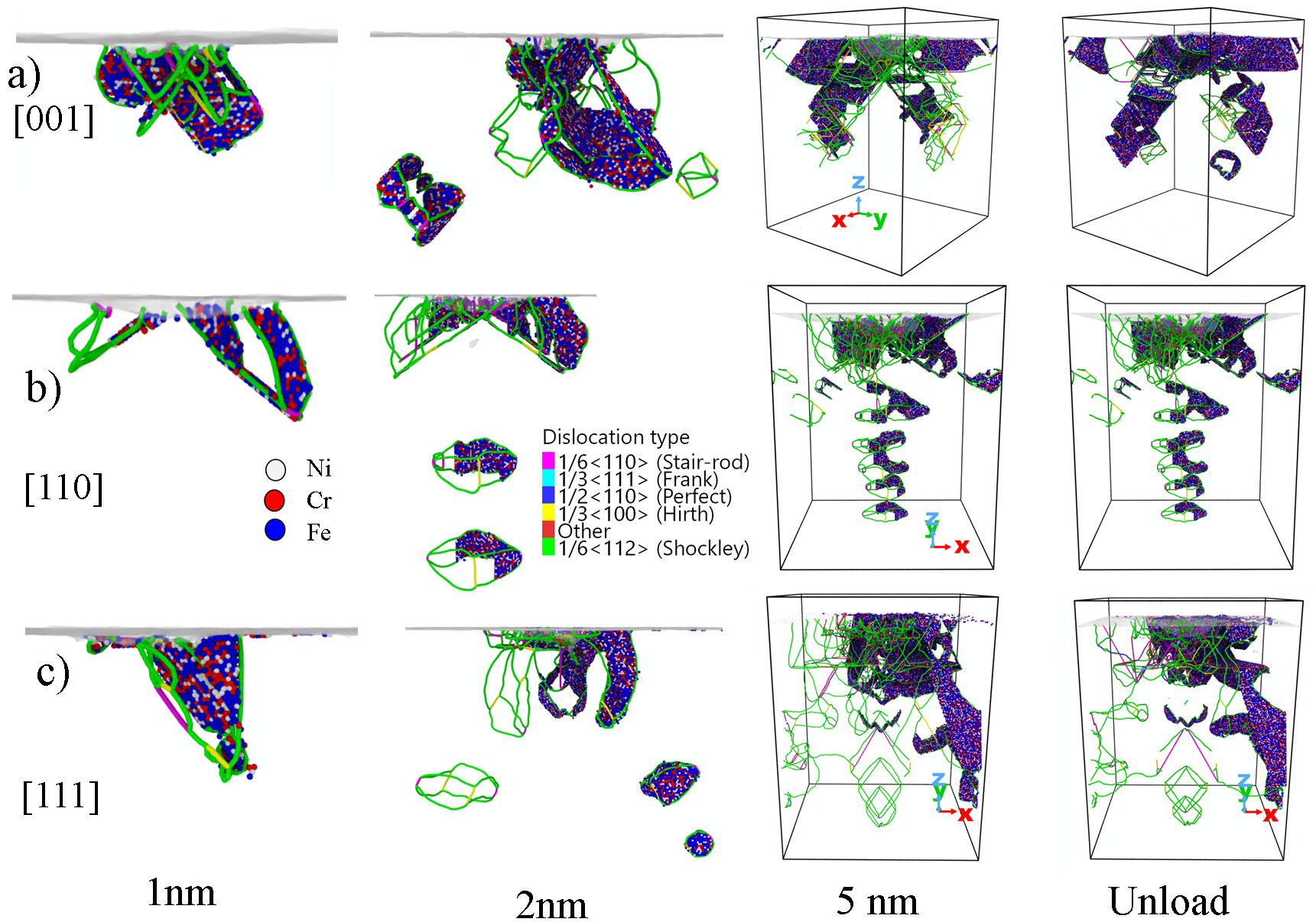}
   \caption{(Color on-line). Dislocation lines and loops nucleated 
   during the loading process at 1nm, 5nm (max. depth) and after
   unloading process for the [001] in a), for [110] in b), and for
   [111] crystal orientation in c). 
   Atoms displaced during the loading and unloading process are
   included to identify the stacking fault \{111\} planes. 
   The dislocation types are colored according to their Burgers 
   vectors as: ½<110> Perfect (blue), 1/6<112> Shockley (green),
   1/6<110> Stair-rod (red), 1/3<100> Hirth (yellow), 1/3<111> Frank (turquoise).}
   \label{fig:fig8}
\end{figure*}

In Fig \ref{fig:fig8}, we show the dislocation lines and loops nucleated at 1nm,
2nm, 5nm depths, and after unloading process. 
The calculations were carried out for a particular MD simulation for [001] in a),
[110] in b), and [111] 
in c) crystal orientations. We noticed that the dislocation loops following the
\{111\} slips systems by computing the atomic displacements where prismatic 
dislocation loops are found due to the interaction of Shockley type dislocation 
during the loading process. Obtained numerical results shows that the mobility 
of prismatic dislocation loops (PDLs) on the [001] crystal orientation tend to form stacking fault tetrahedron 
(SFT) due to the 1/3<100> Hirth dislocation junction by the interaction of Shockley
type dislocations. At the same time, the formation of SFT is not observed for the 
[110] and [111] orientations. 
On the [110] orientation, we observe the propagation of PDL through the sample 
where 4 loops were found at the maximum depth. After the unloading process, 2 PDLs
are pushed back and they are absorbed by the surface, leaving 2 PDL in the sample 
after the nanoindentation test. 
Lastly, for the [111] orientation, 4 PDLs were observed at the maximum indentation
depth. 
Finally, the effect of hardness decrease is connected with the 
nucleation the dislocation loops at the 2 nm depth acting subsequently as obstacles to the
deformation in the material and increase of hardness.
The unloading process provides the information about elastic recovery after
the nanoindentation cycle is terminated. In the Supplementary material of this
paper we show the animation videos of the MD simulations of the loading and
unloading processes at different crystal orientations.

\begin{figure}[t!]
   \centering
   \includegraphics[width=0.45\textwidth]{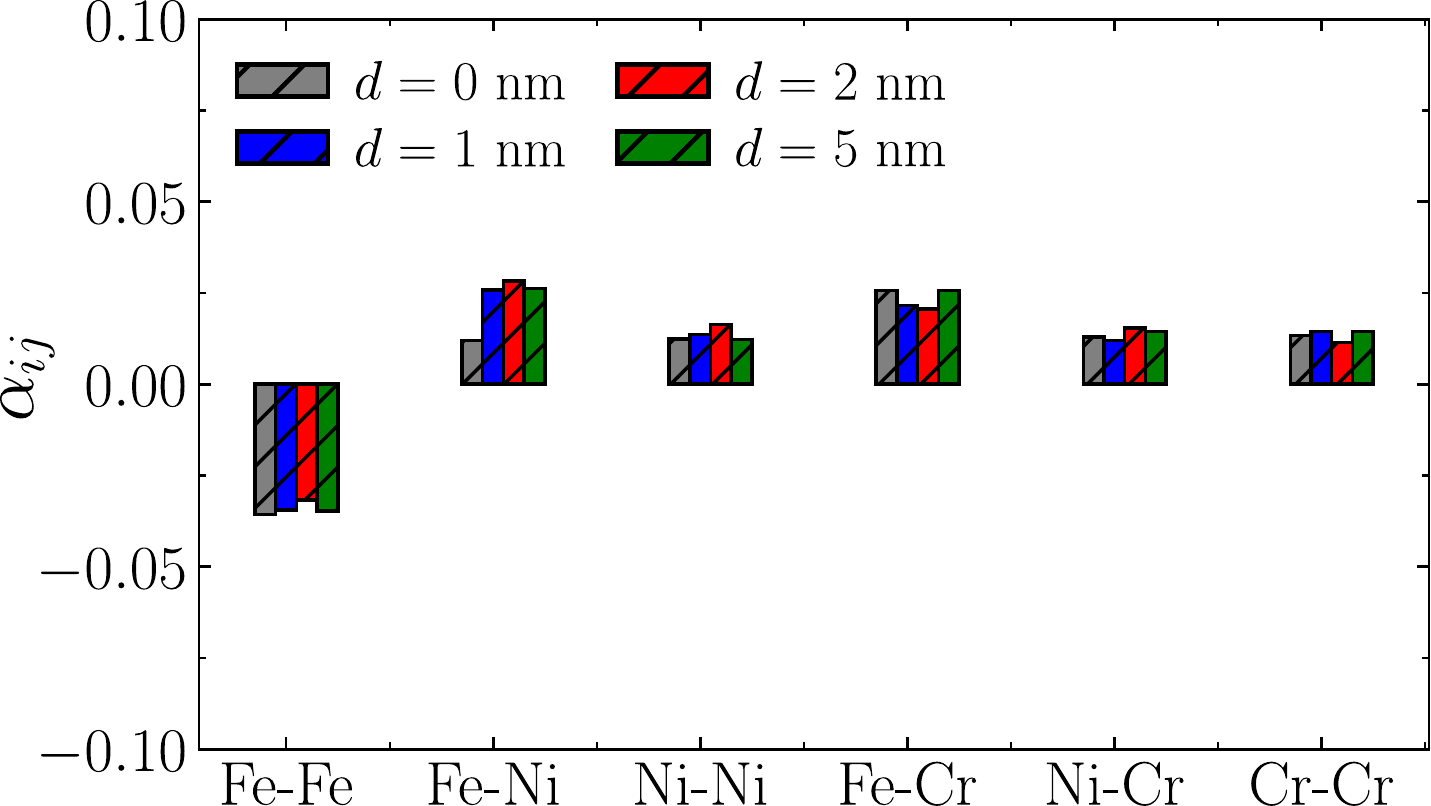}
   \caption{(Color on-line) Pair wise short-range order parameter
   computed at different nanoindentation depths.}
   \label{fig_rene}
\end{figure}

In the next step we checked if the nanoindentation simulation induces a 
level of ordering in our random sample due to the chemical complexity 
of the stacking fault $\{111\}$ planes formed during loading process. 
To do so, we calculate the pair wise shot range order parameter 
as \cite{li2019strengthening}:

\begin{equation}
    \alpha_{ij}^{m=1} = \frac{p_{ij} - C_j}{\delta_{ij}-C_j},
\end{equation}

where $p_{ij}$ is the average probability of finding a $j-$type
atom around an $i$-type, $m=1$ means that the calculation is
performed only considering the 1st nearest neighbors, 
$C_j$ is the average concentration of $j-$type atom in the sample,
and $\delta_{ij}$ is the Kronecker delta function.
Fig \ref{fig_rene} shows the value of $\alpha_{ij}$ at 1, 2, and
5nm depths. 
An evident increase in the absolute value of the pairwise SRO 
parameter is not observed indicating that nanoindentation test
does not affect the randomness of the SS samples; and the 
system keeps as a random solid solution structure during the
whole process. In addition, this fact was also found during the 
nanoindentation test of equatomic NiFe sample.

In nanoindentation testing, plasticity size effects were vastly experimentally studied 
\cite{Yilmaz2020, Ustrzycka2020, Hahner_2018}. In the case of strain-hardening 
materials, hardness is less related to yield strength and plastic deformation, 
but it directly depends on a strain induced by the indenter geometry. 
Here, Nix and Gao \cite{Nix1998} proposed the mathematical model by considering the
total density of dislocations separated into two densities
of statistically stored dislocations (SSD) and geometrically stored dislocations (GND).
The former constitutes a group of dislocations accumulated by multiplication
during plastic deformation and the latter is accumulated in strain gradient fields
caused by geometrical constraints of the crystal lattice. 
%


For the considered material the depth-dependent nanoindentation
hardness is plotted as a function of the indentation depth in
Fig \ref{fig:fig5} noticing a suddenly decrease for both MD simulation and 
experimental method. This occurs at around 80 nm indentation depth from
the experimental measurements and around the 2 nm depth for the
numerical modelling. 
Thus, we note that the decrease of the measured hardness of the material
is observed at early stages of nanoindentation process, where
%
the accumulation rate of GNDs can be described by the geometric slip distance, 
parameter strongly dependent on the microstructure and independent of 
strain \cite{Ashby1970}. 
The geometric slip distance is analogous to the slip distance for
the SSD and expresses the effectiveness of particles or grain in
causing dislocations to be stored. 
Moreover, the GND density has to be directly related to the strain gradient 
as follows:

\begin{equation}
   \rho_{GND} =\frac{4 \gamma}{bl}
    \label{Eq:GND_Ashby}
\end{equation}

where $\gamma$ is the shear applied on the primary slip plane and $l$ is
defined as some finite length connected with the particle size.
For a rough estimate of the geometrically stored dislocations GNDs,
Ma and Clarke \cite{Ma1995} proposed the average shear strain below
the indent as

\begin{equation}
   \rho_{GND} =\frac{4 \gamma_{avg}}{bD}
    \label{Eq:Ma}
\end{equation}
where $D$ denotes the indent diameter.
The local length scale used in Eqs. \ref{Eq:GND_Ashby} and \ref{Eq:Ma} 
can be expressed by the indentation contact depth as $D=h_c$. 

\begin{figure}[t!]
   \centering
   \includegraphics[width=0.475\textwidth]{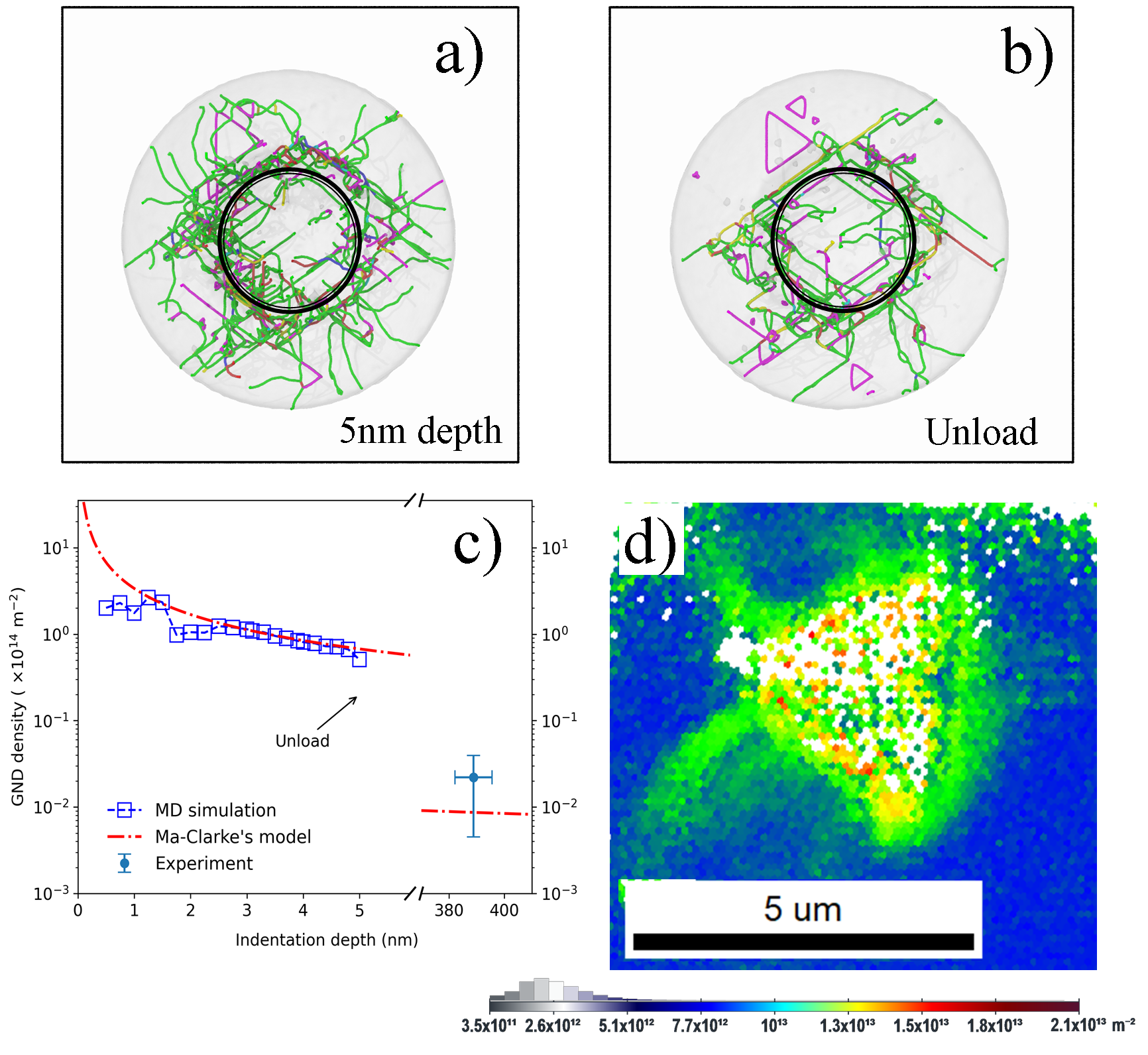}
   \caption{(Color on-line). GND visualization by MD simulation
   for the maximum indentation depth in a) and unloading in b)
   on the [110] orientation; plastic volume is depicted by a
   gray colored sphere. 
   GND density as a function of the depth by MD simulations,
   Ma-Clarke's model, and experimental data in c).
   EBSD image for the GNDs mapping observed on a indented 
   grain with [110] crystal orientation in d).}
   \label{fig:figgnds}
\end{figure}

In Fig \ref{fig:figgnds}a-b) we present the visualization of the 
GNDs nucleated at the maximum indentation depth and after nanoindentation test, 
a gray circle is added the plastic deformation volume.
We noticed that computed GNDs density from MD simulation as a function 
of indentation depth follow the Ma and Clarke relationship as shown in 
Fig \ref{fig:figgnds}c)  
(Fig \ref{fig:figgnds}d) reports experimental data obtained by 
EBSD investigation of the indentation made with depth of 388.68 nm 
(see Fig. \ref{fig:figgnds}d) and a crystal orientation close 
to [110] where the obtained GND density agrees well with the Ma-Clarke's 
fitting. 
Due to the fact, that the plastic area around the indent is 
proportional to the ten times of the indentation depth 
\cite{Matucci2021} the GNDs analysis was provided from the 
larger area (compare the zones marked with red circles in Fig.
\ref{fig:figgnds}d). 
Thus, the surface distribution of GNDs density
is obtained.

\begin{figure}[b!]
   \centering
   \includegraphics[width=0.23\textwidth]{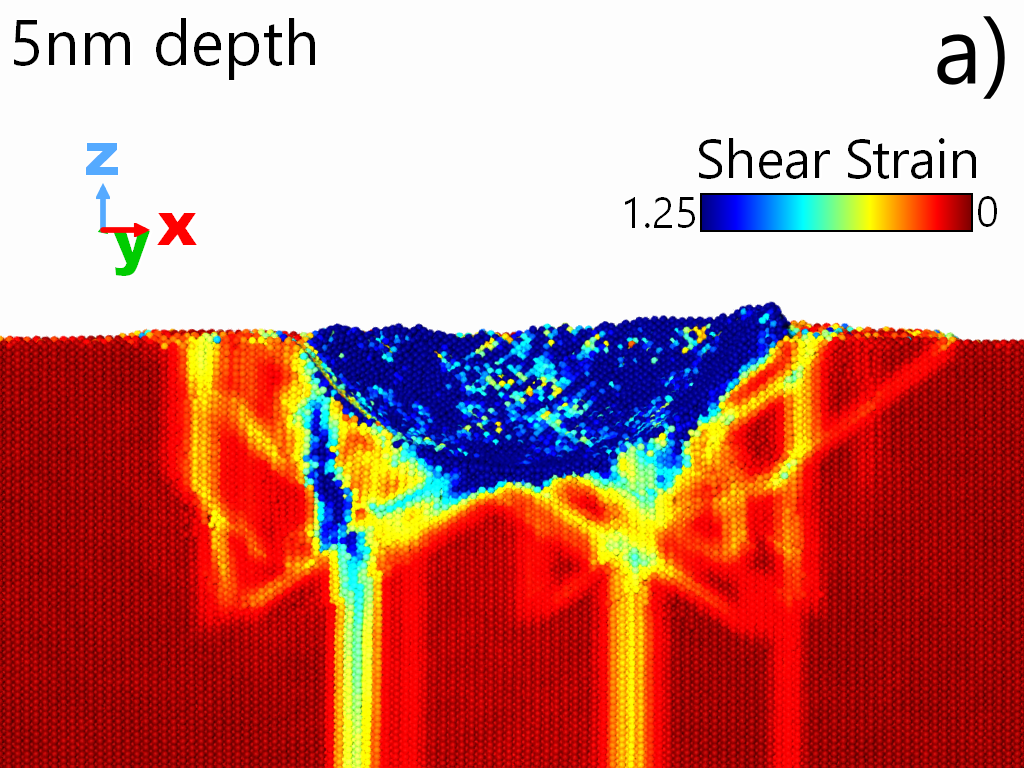}
   \includegraphics[width=0.23\textwidth]{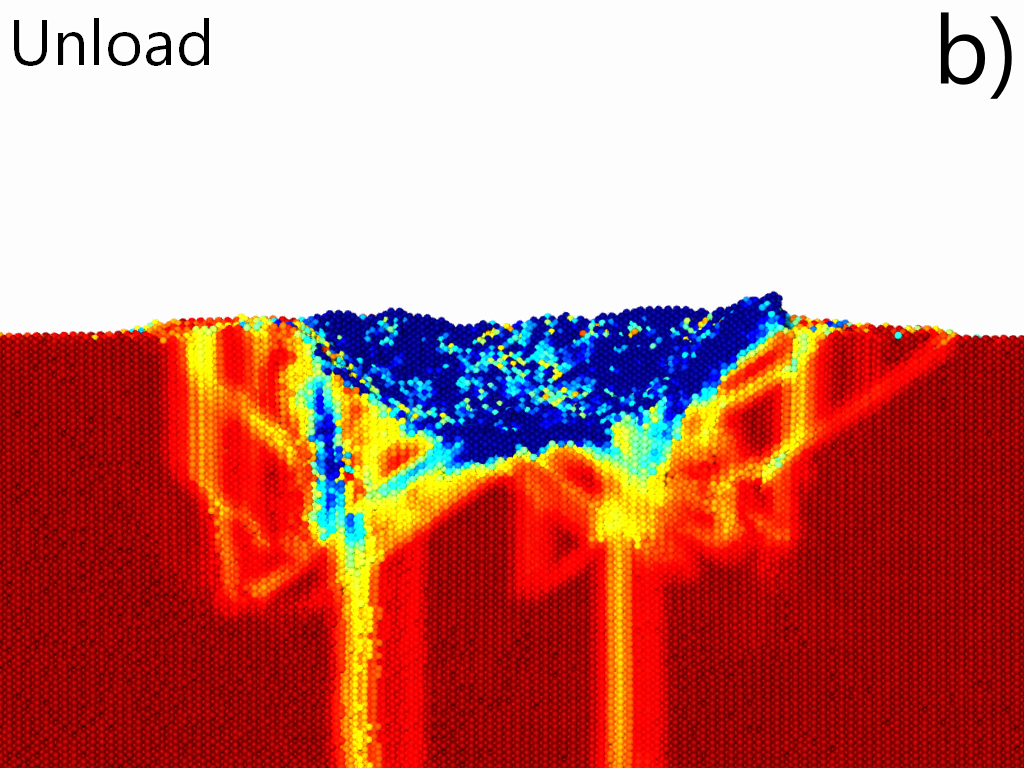}
   \caption{(Color on-line) von Mises strain mapping at 
   maximum indentation depth in a) and after unloading process in b) 
   of the [110] SS sample.}
   \label{fig:strainappendix}
\end{figure}

We have also noted that GNDs created in the volume deformed by
the indentation are directly related to lattice distortion which 
suggests that the production of GNDs has a considerably key role 
on strength of stainless steel. 
In addition, increasing atomic level disorder can lead to reduction 
in mean free paths of dislocation and a significant impact
on evolution of defect formation,
where the presence of multi-elements makes slip paths of 
dislocations differently shaped than for single element metals \cite{Javier2021}. 
We have also noted that the first pop-in event identified in the
nanoindentation tests (Fig \ref{fig:fig4}) demonstrate the onset
of microplasticity and the nucleation of the first dislocations
but the second pop-in event during the loading process is connected
with the drop of the hardness. 
Atomic shear strain for SS samples at [001]
orientation, are shown in Fig \ref{fig:strainappendix}. 
The sample is slid to the half in the \{111\} slip plane 
to visualize the atomic distribution of the shear strain 
by coloring Fe, Ni, and Cr atoms according to their values; results
for the 
plastic region are presented for the maximum indentation depth 
and after unloading process. 
We notice the correspondence between the GNDs nucleated into 
the plastic region and the slip trace formation, where the 
shear strain has values of 0.1; besides that, the maximum values 
are observed underneath the indenter tip at the maximum depth 
and by the in-print mark after unloading process.

\section{Concluding Remarks}
\label{sec:conclusion}

In this work, we present a joint experimental and atomistic based
computational study of nanoindentation mechanisms of  
310S stainless steel. 
We numerically model the dislocation nucleation and evolution in 
different crystal orientations where trend in hardness and young’s modulus 
data between experiments and simulations reached an excellent agreement. 
We discuss the decrease of hardness at indentation depths close
to the surface due to the lattice mismatching of the material 
and its elastic-plastic deformation transition where the formation
of dislocation loops affects the material mechanical properties. 
Thus, we characterized the nanoindentation process in connection
to experimental findings, and through tracking dislocation dynamics
and densities at different indentation depths 
where prismatic dislocation loops are nucleated and mainly formed
by Shockley type dislocations; regardless the particular 
grain orientation and its chemical disorder. 

Several new elements 
can be summarized: 
1) The analysis of the nanoscale anisotropic elastic–plastic behavior 
of SS using nanoindentation, as 
well as pile-up patterns and atomic strains 
distribution using molecular dynamics (MD) simulations; 
2) The elucidation of atomistic mechanisms of dislocation nucleation
and defects evolution by suitable nano-indentation tests in
SS by comparison to equiatomic Ni-Fe solid solutions under identical conditions~\cite{KURPASKA2022110639};
3) The verification of the connection between geometrically necessary 
dislocations (GND) and size effects based on  dislocation
dynamics and associated continuum modeling.
From our results, we can argue that low-carbon austenitic stainless
steel 310S indicates anisotropic properties and
the insights into lattice defect dynamics may provide a basis for 
understanding the physical mechanisms associated with its deformation
making it a good candidate 
for applications in nuclear power plants and extreme operating 
environments.
Our work is also aimed to inspire the design of more advanced chemically 
complex functional materials and understand their mechanical properties
at operating conditions of future advanced nuclear reactors. 

\section*{Acknowledgments}
 We acknowledge support from the European Union Horizon 2020 research
 and innovation program under grant agreement no. 857470 and from the 
 European Regional Development Fund via the Foundation for Polish 
 Science International Research Agenda PLUS program grant 
 No. MAB PLUS/2018/8 (R.A.D.,L.K.,S.P., and M.A.). 
 This work has been partially supported by the National Science Centre through the Grant No UMO-2020/38/E/ST8/00453 (F.J.D.G.,K.M., A.U.).
 We acknowledge the computational resources 
 provided by the High Performance Cluster at the National Centre 
 for Nuclear Research in Poland, as well as 
 the support of the Interdisciplinary Centre for Mathematical and
 Computational Modelling (ICM) University of Warsaw under 
 computational allocation no g88-1181.




\bibliographystyle{apsrev4-1}
\bibliography{references}

\end{document}